\documentstyle[12pt]{article}
\begin{document}
\pagestyle{empty}
\sloppy
\begin{center}
{\Large {\bf Extended states in 1D lattices : application to quasiperiodic
copper-mean chain}}\\
\vskip 2.0cm
S. Sil, S. N. Karmakar, R. K. Moitra\\
{\em Saha Institute of Nuclear Physics\\
1/AF Bidhannagar, Calcutta 700 064, India}\\
and\\
Arunava  Chakrabarti\footnote{{\bf  Permanent  address}  :  Department   of
Physics,
Scottish Church College, 1~\& ~3 Urquhart Square, Calcutta 700006, India}\\
{\em Theory Group, H. H. Wills Physics Laboratory\\
Tyndall Avenue, Bristol, BS8 1TL, U.K.}
\end{center}
\newpage
\vskip 3.0cm
\begin{abstract}

The question of the conditions under which 1D systems support
extended electronic eigenstates is addressed in a very general context.
Using real space renormalisation group arguments we discuss the precise
criteria for determining the entire spertrum of extended eigenstates
and the corresponding eigenfunctions in disordered as well as quasiperiodic
systems.  For  purposes  of  illustration  we  calculate  a  few   selected
eigenvalues
and the corresponding extended eigenfunctions for the quasiperiodic
copper-mean chain. So far, for the infinite copper-mean chain,
only a single energy has been numerically shown to support an extended
eigenstate [ You et al. (1991)] :
we show analytically that there is in fact an infinite number of extended
eigenstates in this lattice which form fragmented minibands.
\end{abstract}
\newpage
\pagestyle{plain}
\pagenumbering{arabic}
     In recent years  extended  electron  states  have  been  found  in   a
variety
of diverse 1D systems [1-14] ranging from  disordered  systems  on
the one hand to various kinds of quasiperiodic systems  on  the  other.  In
most
cases, such states are identified through  extensive numerical
computations [1-8].
 A general  analytical understanding of this problem is still lacking
 [9-14].
In this paper we like to address this problem and  we  wish  to  see  under
what
physical conditions extended states are expected to  occur  in   disordered
or
quasiperiodic systems.

     Consider  an  elementary  example   --   an   Anderson    localisation
problem
involving the random distribution of  A atoms on a  host  B-atom   lattice.
It
is well-known \cite{ref3,ref10} that  there is  an extended eigenstate at a
certain energy
in this system if the A-atoms always occur in pairs, i.e.,  in   the   form
of
dimers. This is an instance of  a  definite  kind  of  correlation  between
atoms
leading to extended states  in  a  disordered   system.   In   this   case,
the
correlation consists  in the  fact that an A atom {\em always\/} occurs  as
a
member
of a dimer. We may generalise this idea as follows.   Suppose  that   there
is
a certain cluster of atoms  which  is  distributed  in  some  manner  on  a
linear
chain. Let this   cluster   consist   of   a   finite   repitition   of   a
certain
subcluster of atoms. Thus in the above example, a cluster  (an   AA   pair)
is
made up by repeating a subcluster (an A  atom)  a   finite   number   (two)
of
times. If this  subcluster is to be  found   only   within   the   clusters
and
nowhere else in the chain, then it  can  be  shown \cite{ref12}  that   for
certain
discrete energy values these clusters   offer  identity   contributions  to
the
total transfer matrix \cite{ref15}.
At these special energies  we may  disregard  the
presence  of  the  clusters  themselves,  and  if  it  happens   that   the
remaining
part of the lattice  is a periodic  one,   then  evidently   this   lattice
will
support extended states at these energies.

          The  special  energies  which  support  extended  states  are   a
property
of the cluster only, and do not depend on the lattice as  a  whole.   Thus,
in
our  example of randomly   distributed  A-type  dimers  on  a  B-type  host
lattice
(RDL), the special energy value $E = \varepsilon_A$ at   which   there   is
an
extended state is determined from  the  consideration  of  the  AA  cluster
alone
\cite{ref12}.
At this energy value these AA dimers make  identity  contributions  to
the total transfer matrix and the  chain  effectively  consists   only   of
B
atoms. We now wish to emphasize that  many  other  1D  lattices  containing
these
AA dimers may have extended  states at  this  same  energy.  We  may
illustrate this by considering  an  ordered  chain  consisting  only  of  A
atoms
and comparing its spectrum with that of the RDL. In the ordered  chain  the
AA
correlation is trivially present, and  therefore   this   system   supports
an
extended state at the same energy. However, and this  the   crucial   point
we
wish to make, while all eigenstates at other energies are localised
ones in the RDL, in the ordered chain of A atoms we have   a whole band
of extended states, the existence of which {\em cannot\/} be inferred  from
a consideration of the AA dimer alone. Although  this  band  can  be  found
directly by using  Bloch's  theorem,   we   do   encounter   many   systems
also
possesing bands or minibands  of  extended  states   [5-14],   where   this
theorem
cannot be used as there is  no  periodicity.  An   example   of   this   is
the
quasiperiodic copper-mean chain (CMC).

          The clusters contain  only part   of   the    information   about
the
entire lattice, and therefore, provide  a  common  set  of  energy
levels for a variety of  systems   which   contain   these   clusters   but
which
otherwise differ  among  themselves  in  their  long  range  behaviour.  In
general,
these  systems possess correlations beyond the level of what  is  contained
in
the  clusters  themselves,  which  are   responsible   for   the   detailed
differences
in   electron  spectra  among   them.  In  order  to  bring    out    these
differences
we have to examine correlations among larger and larger   blocks  of  atoms
in
the chain, by looking at the system at larger and  larger  length   scales.
In
this process the  long  range  features   of   the   lattice,   which   may
include
periodicity, or may  be  self-similarity,   get   gradually   included   in
the
determination of the electronic spectra, and  we  end   up   by   obtaining
the
entire spectrum of  extended  states.

          In this paper we present  an  approach  which  incorporates  this
point
of view for finding the full spectrum of extended states in  1D
lattices, periodic, quasiperiodic or disordered.  The most  natural  method
of
including the effects of correlations at  all   length   scales    is   the
real
space renormalisation group (RSRG) method, which is adopted here.

     Although the discussion so far has been perfectly  general,  from  now
on
we illustrate our ideas by focussing attention on the CMC. A portion  of
a CMC is shown in Fig.~1. The sequence  in  which  long  ($L$)  and   short
($S$)
bonds are arranged in a CMC may   be   obtained   from   the   substitution
rule
$L\rightarrow LSS$ and $S  \rightarrow  L$   starting   with,   say,   $L$.
For
decribing the electron  states  in  this   lattice   we   use   the   tight
binding
hamiltonian
\begin{equation}
              H = \sum_{i} \varepsilon_{i} |i\rangle \langle i| +
\sum_{<ij>} t_{ij} |i\rangle \langle j| \label{eq1}
\end{equation}

          As shown in Ref. \cite{ref7}, for implementing the RSRG method in
CMC  we
have to identify  four types of sites, namely,  $\alpha$, $\beta$,
$\gamma$  and
$\delta$  corresponding respectively to the   $L-L$, $L-S$, $S-L$
and   $S-S$
vertices  (Fig.~1).  The  site  energies  in  Eqn.(\ref{eq1})  assume  the
values
$\varepsilon_{\alpha}$,   $\varepsilon_{\beta}$,   $\varepsilon_{\gamma}$
and
$\varepsilon_{\delta}$,  and  there  are  two  different  values   of   the
nearest
neighbour hopping integrals,  $t_{L}$ and $t_{S}$.

          The eigenfunctions for such an 1D lattice may  be
          calculated  by  the
standard transfer-matrix method \cite{ref15}. The discretized Schrodinger
equation can be cast in
the form $\phi_{n+1} = M_{n} \phi_{n}$, where
\[  \phi_{n}  =   \left(   \begin{array}{c}    \psi_{n}    \\    \psi_{n-1}
\end{array}
\right) \mbox{\rm and } M_{n} = \left( \begin{array}{cc}
\frac{E-\varepsilon_{n}}{t_{n,n+1}} & - \frac{t_{n,n-1}}{t_{n,n+1}} \\
1 & 0 \end{array} \right). \]
Here $\psi_{n}$ denotes the amplitude of the wavefunction at the n-th site
and $M_{n}$ is a $2\times 2$ transfer matrix. In the CMC,
$\phi_{n}$ is related
to  $\phi_{0}$  by  a
product of four types of transfer matrices $M_{\alpha}$, $M_{\beta}$,
$M_{\gamma}$
and $M_{\delta}$ following the copper mean sequence.

          By inspecting the CMC (Fig.~1) we  see  that  the  $\alpha$-sites
{\em
always\/} occur in pairs,   whereas,   $\beta$,   $\gamma$   and   $\delta$
{\em
always\/} form a triplet $\beta\delta\gamma$.  Thus,  for   this   lattice,
the
pair  of  sites  $\alpha\alpha$  constitutes  a  cluster.   If    we    now
renormalise
the lattice by  applying    the  deflation  rule \cite{ref7}
$LSS  \rightarrow  L$  and
$L\rightarrow S$, we get a scaled version  of  the  original  lattice  with
scale
factor $\sigma = 2$.  In the renormalised  chain we again find  $\alpha$
sites  occuring  as  $\alpha\alpha$  pairs,  and  $\beta$,  $\gamma$    and
$\delta$
sites forming $\beta\delta\gamma$ triplets. This means that the
$\alpha\alpha$ clustering effects  are  also
present on  this  inflated  length  scale,  and  therefore,  by  induction,
present
at  all  length  scales.  It  is   important   to   appreciate   that   the
$\alpha\alpha$
clustering at larger length scales amounts to   subsuming,   so   to   say,
the
effects of larger and  larger  segments  of   the   original   chain   into
the
$\alpha$ subclusters, thereby allowing larger  and   larger   portions   of
the
chain to contribute to the special energies for the extended states.

          It  is   clear   from   the    above    discussion    that    the
$\alpha\alpha$
correlation  at  all  length  scales can  only  be   revealed   by
renormalisation  group  methods.  While  in  the  original   lattice,
this
correlation is directly visible (Fig.~1), higher  order  correlations   due
to
renormalised   $\alpha\alpha$     pairs    imply     underlying     complex
correlation
between atoms, which is not apparent from mere inspection of the original
lattice.

          In  a  copper-mean  lattice  the  string  of   transfer  matrices
typically looks like,
\begin{equation}
\cdots M_{\beta\delta\gamma} M_{\alpha}^{2}
M_{\beta\delta\gamma}^{3} M_{\alpha}^{2} M_{\beta\delta\gamma} \cdots,
\label{eq2}
\end{equation}
where, $M_{\beta\delta\gamma} = M_{\gamma} M_{\delta} M_{\beta}$.

          We notice that the matrix $M_{\alpha}$ is  unimodular,   and   so
we
can apply a well-known result, due originally to Cayley and Hamilton,
for the \mbox{m-th} power of a  $2\times  2$
unimodular matrix M :
\begin{equation}
M^{m} = U_{m-1}(x)M - U_{m-2}(x)I
\label{eq3}
\end{equation}
where, $x =(1/2) {\rm Tr} M$. $U_{m}(x) = \sin (m+1)\theta/\sin  \theta$,
with $\theta=\cos^{-1}x$ is the m-th Chebyshev polynomial  of  the
second kind. Let  the  matrix  $M^{m}$  correspond  to  the
transfer matrix of a cluster  composed  of   $m$   identical   subclusters,
each
described by a $2\times 2$ unimodular transfer matrix $M$.  For   a   value
of
the energy $E$ for  which  $U_{m-1}(x)$   becomes   equal   to   zero,   we
have
$M^{m}= -U_{m-2}(x)I$, that  is  to  say,  the  transfer  matrix  for  this
cluster
behaves essentially as the identity matrix  at   this   energy.   Thus   at
this
energy, the  lattice does not `feel' the presence of the  clusters  defined
by
the transfer matrix $M^{m}$. If the   remainder  of   the   lattice   forms
a
periodic chain, there will be an extended state  at  this  energy  provided
this
energy  is an  allowed  one.  For  allowed  states  wavefunctions  do   not
diverge
at infinity \cite{ref15}.

          For the CMC, therefore, we  have to put $m=2$ in Eqn.(\ref{eq3}).
          We  see
that $U_{1}(x) = 0$ for $E  =  \varepsilon_{\alpha}$,  and  therefore,  for
this
energy $M_{\alpha}^{2}=-I$ as $U_{0}(x) = 1$.  Thus,   for   this   energy,
the
string of  transfer  matrices  given  by  Eqn.~(2)  is  composed  only   of
matrices
$M_{\beta\delta\gamma}$.  This   effectively   describes    an    `ordered'
linear
chain composed of  unit  cells  each  containing  three  types   of   atoms
$\beta$,
$\delta$ and $\gamma$. Now, if the condition
$(1/2)|{\rm Tr} M_{\beta\delta\gamma}| \leq 1$
is satisfied for  $E  =  \varepsilon_{\alpha}$,  then  this  energy  is  an
allowed
one (see Ref.\cite{ref15}), and we shall have an  extended  eigenstate  for
the  whole
system.

          To  determine  all  the  other  extended  states,  let   us   now
consider
all  the  successive  renormalised  versions   of   the   original   chain.
The
self-similarity  of  the  lattice  implies   that   one   can   apply   the
considerations
discussed so far in this paper to every  such  renormalised   version.   At
any
stage $\ell$ of renormalisation, extended states will be found at energies
for    which    $E    =    \varepsilon_{\alpha}^{({\ell})}$,     where
$\varepsilon_{\alpha}^{({\ell})}(E)$ is the  renormalised  site-energy  of
$\alpha$      site      at      this      stage.       The       expression
for
$\varepsilon_{\alpha}^{({\ell})}$  can  be  obtained  by  iterating   the
recursion relations \cite{ref7}
\begin{eqnarray}
\varepsilon_{\alpha}'=\varepsilon_{\gamma} + Q(t_{L}^{2}P_{\beta} +
t_{S}^{2}P_{\delta}), &
\varepsilon_{\beta}'=\varepsilon_{\gamma} + Qt_{S}^{2}P_{\delta}, &
\varepsilon_{\gamma}'=\varepsilon_{\alpha} + Qt_{L}^{2}P_{\beta},
\nonumber \\
\varepsilon_{\delta}'=\varepsilon_{\alpha}, &
t_{L}'=Qt_{S}^{2}t_{L}P_{\beta}P_{\delta}, & t_{S}' = t_{L},
\label{eq5}
\end{eqnarray}
where              $Q=(1-t_{S}^{2}P_{\beta}P_{\delta}               )^{-1}$
and
$P_{i}=(E-\varepsilon_{i})^{-1}$, with $i  =  \alpha  ,  \beta   ,   \gamma
,
\delta$.

     The roots of the polynomial $E-\varepsilon_{\alpha}^{({\ell})}$  for
every ${\ell}$ yield a new set of energy values for which eigenstates  are
extended. The number of allowed energy values increases with  the  progress
of
iteration. The   totality   of   all   these   energies   constitutes   the
entire
spectrum of extended states for the CMC. These  energies   form   minibands
in
energy space,  the  existence  of  which  was  only   detected   previously
through
numerical work \cite{ref7}.

          The   CMC    may    be    divided    into    three    sublattices
$\Omega_{1}$,
$\Omega_{2}$ and $\Omega_{3}$ (Fig.~1), each being a  scaled   version   of
the
original lattice \cite{ref7}. The recursion relations in Eqn.(\ref{eq5})
actually represent
renormalisation of the original chain to the  $\Omega_{1}$  sublattice.  We
can
define  similar   RSRG   transformations   for   the    $\Omega_{2}$    and
$\Omega_{3}$
sublattices.  Let  us  denote  the   transformations   for   these    three
sublattices
by $T_{1}$, $T_{2}$ and  $T_{3}$. Previously     we     have     determined
$\varepsilon_{\alpha}^{({\ell})}$  by  applying  $T_{1}$ transformation
${\ell}$
times
successively.  In  general  $\varepsilon_{\alpha}^{({\ell})}$  is  to  be
determined by applying a string of ${\ell}$ operators consisting of
$T_{1}$, $T_{2}$  and $T_{3}$  in
any arbitrary sequence.

     A string of $T$'s represents a genealogical path in the sense  of
     Ref.\cite{ref16},
and all the extended eigenstates corresponding to a given string  of  $T$'s
can therefore be considered  as  members  of  the  same   `family'.   These
states
are characterised by the same length scale,  and   they   belong   to   the
same
point on  the  genealogical  tree.  Now  it  may  happen  that  the  states
belonging
to the same family appear in different regions of the  energy  spectrum,
that is, they  may  belong  to  different  minibands.   In   other   words,
two
neighbouring states belonging to a small   energy   interval   $\Delta   E$
may
belong to entirely different families. This indicates  that  the  minibands
are
composed   of    extended     states     which     are,     in     general,
characteristically
different from each other, forming dense sets of discrete energies.

          We now present some results   of   our   calculations   for   the
CMC.
Taking   the    hamiltonian     parameters     as     $\varepsilon_{\alpha}
=
\varepsilon_{\beta} = \varepsilon_{\gamma}  =  \varepsilon_{\delta}  =  0$,
and
$t_{S}/t_{L} = 2$,  the  first  few   stages   of   renormalisation   yield
the
extended state energy eigenvalues
 $E^{(0)}=0$, $E^{(1)}=\pm 3$, $E^{(2)}=\pm 1.3637980258 {\rm  ~and}
\pm 3.0232523786$, and so on.  Only the case $E^{(0)}= 0$
has been detected earlier \cite{ref8} through numerical calculations.
In  Fig.~2  we  have  displayed  wavefunctions  for   a     20{\em    th\/}
generation
copper-mean lattice consisting of  699051  bonds.  The  first   column   of
Fig.~2
shows the amplitudes on the  first   450   sites   of   this   chain,   for
three
selected energy values belonging  to  three  different   genealogies.   The
next
column in this figure shows the amplitudes on  the  last   450   sites   on
this
chain, for the same three  energy  values.  Examining  across  a   row   we
notice
that the `local pattern' of the wavefunction   is   the   same   throughout
the
chain, with the  amplitudes  oscillating  in  an  aperiodic  manner  within
definite
bounds without decaying, unlike the behaviour displayed by a critical or
localised eigenfunction.
The pattern itself is different  for  the  three  different
rows. Although these states are extended, they are not Bloch-like. For $E =
0$  Gumbs and Ali \cite{ref17}
found that the amplitudes decay with a power law.
 However our analysis and explicit
evaluation of the  wavefunction  clearly  shows   that   their   conclusion
is
incorrect.

        As has been shown in an earlier work \cite{ref11},
 in a quasiperiodic lattice the flow pattern
for the hamiltonian parameters under renormalisation at energy eigenvalues
for extended
states displays a regularity which is absent in the non-extended  case.  In
the
CMC, this regularity associated with an extended state shows up in the site
energies for the $\beta$ and $\gamma$ sites becoming equal, after a certain
step of iteration, the equality being maintained in all  subsequent  steps.
The
level at  which the $\beta$ and $\gamma$ site energies  become  equal   for
the
first  time depends on the energy value. Furthermore, this level
is  the  same  for  all energy eigenvalues belonging to the same family.

          The ideas discussed in this  paper  are  very  general. Thus
an  ordered   chain   of   A   atoms   ($\varepsilon_{i}=\varepsilon$   and
$t_{ij}=t$)

may  be  regarded  as
consisting entirely of AA dimers, which remains true at all subsequent
length scales upon renormalisation of the chain. At any level, say $n$,
of iteration the energy
eigenvalues are given by the following analytical expression
\begin{equation}
E^{(n)} =\varepsilon \pm t\sqrt{2\pm \sqrt{2\pm \sqrt{2\pm \cdots
\mbox{\rm n times}}}}  . \label{eq6}
\end{equation}
The eigenvalues $E^{(0)}$, $E^{(1)}$, $E^{(2)}$, \ldots $E^{(\infty)}$
taken together constitute the entire band ranging from
$\varepsilon-2t$ to
$\varepsilon+2t$.
The
same approach applies to the binary ordered chain  of  A   and   B   atoms.
Here
again, the band edges and the band gap  extending  from   $\varepsilon_{A}$
to
$\varepsilon_{B}$ due to the level   repulsion   are   correctly   located.
The
problem of a random  distribution  of  dimers  on  a   lattice   has   been
already
discussed, there being a single energy at which there is an extended
state. This is related to the fact  that  there  is  no  deeper
level of correlation in this  chain  than   what   is   contained   through
the
existence  of  dimers.  If  a  disordered  chain   possess   higher   order
correlations
among the sites in the sense   discussed   in   this   paper   then   there
exist
extended state at  several  energies.  This  may  explain  the various
observations
reported in the literature on disordered systems [1-4].

\newpage
\noindent {\bf Figure captions :}
\begin{itemize}
\item[{\bf Fig.1}] The $\alpha\alpha$ and $\beta\delta\gamma$ blocks in
the copper-mean chain and also in its sublattices.
\item[{\bf Fig.2}] The amplitudes of the eigenfunction on the first
450 sites and the last 450 sites of a 20{\em th\/} generation CMC having
699051 bonds. Here
$\varepsilon_{\alpha} = \varepsilon_{\beta} =\varepsilon_{\gamma} =
\varepsilon_{\delta} = 0$ and $t_{S}/t_{L} = 2$. The 1st, 2nd and 3rd
rows correspond to $E=0$, $-1.3637980258$ and $-2.6255768195$ respectively.
All energies are measured in units of $t_{L}$. We choose the boundary
condition as $\psi_{0}=0$ and $\psi_{1}=1$.
\end{itemize}
\end{document}